\documentclass[treatise,allpages]{ouvrage-hermes}

\usepackage[latin1]{inputenc}
\usepackage[T1]{fontenc}
\usepackage{lscape}
\usepackage{diagbox}
\usepackage{subfigure} 
\usepackage{array,multirow}
\usepackage{url}
\usepackage{orcidlink}
\usepackage[intoc]{nomencl}
\usepackage{natbib}

\usepackage[english,french]{babel}

\setcitestyle{citesep={~;}}
\setcitestyle{aysep={}}
\let\cite=\citep

\definecolor{notecolor}{rgb}{1.0, 0.1, 0.1}

\newcommand{\aap}{A\&A}

\newcommand{\apjl}{ApJL}

\newcommand{\nat}{Nature}

\title{GRAVITATIONAL WAVES}

\makenomenclature 
\makeindex

\begin{document} 



\mainmatter


\newcommand{\hu}{\xspace \ensuremath{{\rm km \, s^{-1} \, Mpc^{-1}}\xspace}}

\ChapterAuthor{GRAVITATIONAL WAVES}{Gravitational Wave Cosmology: an introduction}{Simone \textsc{Mastrogiovanni} and Gregoire \textsc{Pierra}}

\authorname{Gregoire \textsc{Pierra}\textsuperscript{1}\orcidlink{0000-0003-3970-7970} and Simone \textsc{Mastrogiovanni}\textsuperscript{1}\orcidlink{0000-0003-1606-4183}}{\textsuperscript{1}INFN, Sezione di Roma, I-00185 Roma, Italy}


\enlargethispage{2.5pc}

\section{Introduction}

Gravitational-wave (GW) observations have opened a new era in cosmology by providing an independent way to study the cosmic expansion. Compact binary coalescences (CBCs) -- the mergers of black holes (BHs) and neutron stars (NSs) -- emit GWs that carry direct information about their luminosity distance. When combined with information about the redshift of the source, these events can serve as \textit{standard sirens}, providing a new means to measure cosmological parameters such as the Hubble constant, the matter density, and the dark energy equation of state. This method is fundamentally different from traditional EM(EM) techniques, and therefore it is a valuable addition to the tools that we have to study the cosmos.

In this chapter, we develop the necessary tools to understand and implement GW-based cosmological measurements. We aim to provide a soft introduction to all the tools required for GW cosmology.
We begin by summarizing some basic aspects of CBCs and GWs at cosmological distances in Sec.~\ref{sec:1}. In Sec.~\ref{sec:2} we will be setting the core framework of hierarchical Bayesian inference (HBI), a statistical method used to infer population and cosmological properties of CBCs. This approach allows us to infer both the intrinsic properties of individual events and the underlying distributions governing the population, while carefully accounting for selection biases introduced by detector sensitivity. We will show how the merger rate of CBCs is formulated and how cosmological information naturally enters through the relationship between the detector-frame and source-frame parameters.
Henceforth, this chapter proceeds with Secs.~\ref{sec:3}-\ref{sec:5}, in which we will delve into the details of GW cosmology for sources observed with an EM counterpart (Bright sirens) and without an EM counterpart (dark sirens). We will also discuss how it is possible to apply these methods to currently public gravitational-wave data products, while referencing external resources for a more involved reading on the current results.

With this chapter, our goal is to provide a comprehensive and accessible road map to GW cosmology, emphasizing the conceptual flow from the fundamental GW measurements to their applications in constraining the expansion history of the Universe.

\section{Background for standard sirens cosmology}
\label{sec:1}

Direct measurements of the cosmic expansion exploit the observation of sources for which their distances and the recessional velocities can be measured. 
CBCs at cosmological scales are unique; they allow us to directly measure the luminosity distance $d_L$ through their GW signal emitted. Unlike standard candles, which require a known intrinsic luminosity to infer distance, CBCs are self-calibrating sources, earning them the name \textit{standard sirens}. This property allows us to bypass the traditional cosmological distance ladder for cosmology purposes. However, to fully exploit CBCs for measuring the cosmic expansion, the redshift $z$ of the source must also be known to establish the distance--redshift relation. For a more detailed discussion of the theoretical and experimental aspects of GW sources, we refer the interested reader to \cite{Maggiore:2007ulw,Sathyaprakash:2009xs}

In this section, we will derive the expression for the luminosity distance $ d_L(z)$ as a function of redshift, essential for GW cosmology. To do so, we start by revisiting the cosmological framework of an expanding Universe, which is governed by the Friedmann--Lemaitre--Robertson--Walker (FLRW) metric and described by a scale factor $a(t)$. This geometric framework provides the foundation for understanding distances in a Universe that is not static but evolving over time. By relating the observed flux from a source to its intrinsic luminosity and incorporating the effects of cosmic expansion, we will ultimately arrive at the key expression for $d_L(z)$, which ties the observed distance to the redshift of the source. This equation is pivotal for utilizing CBCS as standard sirens and measuring the expansion of the Universe with GWs. We recommend \cite{Dodelson:2003ft}
 for a more dedicated introduction for the field of cosmology.
 
\subsection{An introduction to the cosmic expansion}

Modern cosmology rests on the assumption of the cosmological principle, which states that the Universe is both \textit{homogeneous} and \textit{isotropic} when viewed on sufficiently large scales. Homogeneity means that the Universe has the same properties at every point, while isotropy implies that it looks the same in every direction. Although the Universe exhibits rich structures such as stars, galaxies, and clusters on small scales, observations show that on sufficiently large scales, the matter distribution becomes statistically uniform. A major observational confirmation of this principle is the expansion of the Universe. In the 1920s, Edwin Hubble and Georges Lemaitre independently discovered that distant galaxies are receding from us, with a recession velocity that increases with distance. This relationship is known as the Hubble--Lemaitre law:
\begin{equation}
    v = H_0 d,
    \label{eq:hubble law}
\end{equation}
where $v$ is the recession velocity, $d$ is the distance to the galaxy, and $H_0$ is the Hubble constant, quantifying the present-day expansion rate of the Universe in $\rm km\,s^{-1}\,Mpc^{-1}$. This quantity plays a central role in cosmology and will be discussed further in this section. One of the major open tensions in cosmology concerns the value of the Hubble constant. Observations of the Cosmic Microwave Background yield a value of $H_0 = 67.49 \pm 0.53~\mathrm{km,s^{-1},Mpc^{-1}}$--representing an early--Universe measurement \citep{planckcolab}--whereas local measurements based on Cepheid--calibrated Type Ia supernovae indicate a higher value of $H_0 = 73.04 \pm 1.04~\mathrm{km,s^{-1},Mpc^{-1}}$ \citep{Riess2022}.

The Hubble--Lemaitre law in Eq.~\ref{eq:hubble law} is in fact a product of Einstein's General Relativity when supplied with a homogeneous and isotropic metric, the Friedmann--Lemaitre--Robertson--Walker (FLRW) metric. This metric reads:
\begin{equation}
    ds^2 = -c^2 dt^2 + a^2(t) \left[ \frac{dr^2}{1-k r^2}+r^2(d\theta^2+\sin^2 \theta d\phi^2) \right],
    \label{eq:flrw metric}
\end{equation}
where $ds^2$ is the spacetime interval, $c$ is the speed of light, $a(t)$ is the scale factor that evolves with cosmic time $t$, $dr^2$ the radial component of the metric, often called comoving coordinate, $k$ the curvature of the Universe and $\theta,\phi$ the usual angles for polar coordinates. As we will see later, the scale factor $a(t)$ is responsible for the expansion of the Universe, while the comoving coordinate $dr^2$ represents only the fixed distances defined on the spacetime metric.
Let us now restrict ourselves to the case of a flat Universe ($k = 0$), as supported by most cosmological observations (see \cite{Turner:2022gvw} for a review). We can then center our reference frame on the observer and consider a galaxy emitting photons at some comoving distance $r$, with the condition $ds^2 = 0$. In other words, we are now working with a simplified version of the FLRW metric.
\begin{equation}
    0 = -c^2 dt^2 + a^2(t) dr^2,
    \label{eq:flrw metric 2}
\end{equation}
from which we can learn some interesting properties of cosmological sources.

\begin{itemize}
    \item \textbf{Cosmological sources have a recessional velocity due to the Universe expansion:} By integrating Eq.~\ref{eq:flrw metric}, we can show that the physical distance traveled by the photon is
\begin{equation}
    d_p(t) \equiv \int c dt = a(t) \int dr =  a(t) d_c,
    \label{eq:physical distance}
\end{equation}
where $d_p(t)$ is the physical distance, $a(t)$ is the scale factor, and $d_c$ is the comoving distance, which remains fixed for objects moving with the expansion of the Universe (i.e., with zero peculiar velocity). The scale factor $a(t)$ describes the relative expansion of the Universe over time, representing how distances between objects increase as the Universe expands. Fig.~\ref{fig:scale} provides a visual representation of the interplay between scale factor, comoving distance and physical distance. Since the Universe does not change its scales on the human lifespan, for simplicity, the scale factor is set at $1$ today. In this way, comoving and physical distance coincide.
\begin{figure}
    \centering
    \includegraphics[width=1.0\linewidth]{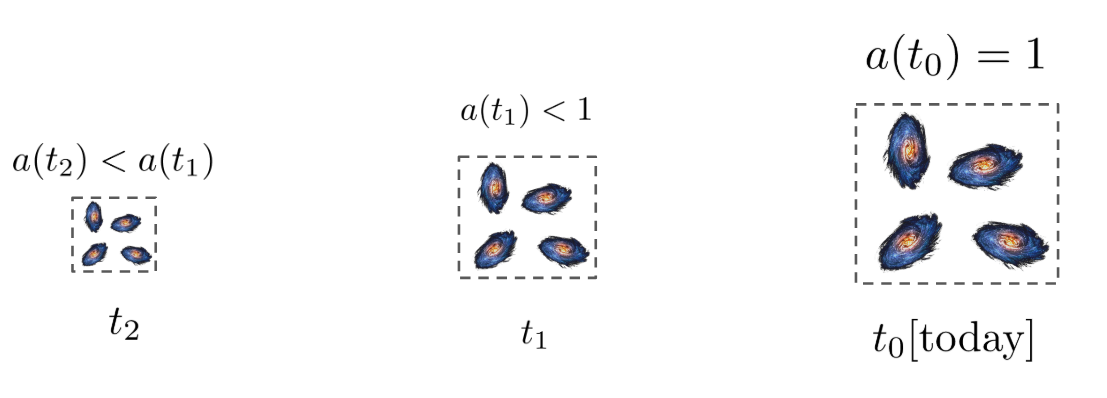}
    \caption{Visual representation of the interplay between scale factor, comoving distance and physical distance. The comoving distance is the distance identified by the coordinates of the square vertexes, the physical distance is the length of the line connecting two vertexes, and the scale factor is how much the line stretches over time.}
    \label{fig:scale}
\end{figure}

Differentiating Eq.~\eqref{eq:physical distance} with respect to time gives back the recession velocity with respect to an observer, which can be expressed as:
\begin{eqnarray}
    \dot{d}_p(t) &=& \dot{a}(t) d_c + a(t) \dot{d}_c, \label{eq: hubble law 2} \\
    &\approx& \frac{\dot{a}(t)}{a(t)} a(t) d_c, \nonumber \\
    &=& H(t) d_p(t), \nonumber
\end{eqnarray}
where we define the Hubble parameter $H(t)$ as the ratio between the time derivative of the scale factor and the scale factor itself:
\begin{equation}
    H(t) \equiv \frac{\dot{a}(t)}{a(t)},
    \label{eq:hubble def}
\end{equation}
and $\dot{d}_c$ the proper motion of the galaxy. The proper motion of galaxies is usually much smaller $\dot{d}_c \ll H(t) d_p(t)$ for far away galaxies, but it might be important to take into account for close by galaxies. For the rest of the chapter, we will assume that this term is negligible. 
In Eq.~\ref{eq:hubble def}, the Hubble parameter $H(t)$ describes the rate of expansion of the Universe at any cosmic time $t$. At the present time $t_0$, or equivalently very close to us, we define $H(t_0) = H_0$ (the Hubble constant), and Eq.~\eqref{eq: hubble law 2} becomes
\begin{equation}
    \dot{d}_p(t)=H_0 d_{p}(t).
    \label{eq:hubble law 3}
\end{equation}
We emphasize that Eq.~\eqref{eq:hubble law 3} is equivalent to Eq.~\eqref{eq:hubble law}. This equation is telling us that we expect cosmological objects (such as galaxies) to present a recessional velocity from us that is due to the expansion. For close by objects, the Hubble constant can be simply measured by obtaining the physical distance of cosmological sources and measuring their recessional velocity. The recessional velocity is ``trivial'' to measure for photons as it can be measured from the redshift of light of known elements, on the other hand, the distance of sources a bit less, as we will see later in the chapter.

\item \textbf{Cosmological sources are redshifted:} As we briefly anticipated in the previous point, since cosmological sources have a recessional velocity from us, they are redshifted. We will now see how the redshift is connected to the scale factor and the definition of the distance.

Imagine the same cosmological source at a comoving distance $r$ emits two light pulses, one at time $t_s$ and the other at time $t_s + dt_s$. These two signals arrive on Earth at times $t_d$ and $t_d + dt_d$, after traveling through an expanding Universe. As the position of the galaxy (source) does not change during this small time interval, we can use Eq.~\ref{eq:flrw metric 2} to equate
\begin{eqnarray}
    \int_{t_s+dt_s}^{t_d+dt_d} \frac{dt'}{a(t')} &=& \int_{t_s}^{t_d} \frac{dt'}{a(t')} \\
    \int^{t_s+dt_s}_{t_s} \frac{dt'}{a(t')} &=& \int_{t_d}^{t_d+dt_d} \frac{dt'}{a(t')},
\end{eqnarray}
and if we assume that the scale factor does not change between the two light emissions\footnote{This is a reasonable assumption, as light pulses are generated by physical processes occurring on timescales significantly smaller than cosmic times}, we can write
\begin{eqnarray}
    \frac{dt_s}{a(t_s)} &=& \frac{dt_d}{a(t_d)} \\
    dt_d &=& \frac{a(t_d)}{a(t_s)} dt_s. 
\end{eqnarray}
The equation above means that two photons emitted with a time interval $dt_s$ are detected with a delayed time interval $dt_d$ that is proportional to the ratio of the scale factors at the epochs of detection and emission. If the detection happens today, with $a(t_d) = 1$ and $a(t_s) < 1$, then their ratio is greater than one, and the photons are redshifted. It follows that we can define a relation between the scale factor and the observed redshift $z$ as
\begin{equation}
    a(t_s) = \frac{1}{1+z}, 
    \label{eq:defs}
\end{equation}
and similarly, signals emitted at a given frequency $f_s$ are observed on Earth at a redshifted frequency 
\begin{equation}
    f_d = \frac{f_s}{1+z}.
\end{equation}

To conclude our discussion on the redshift, let us argue that in cosmology, the redshift is often used as an equivalent of time and distance measures. To understand why, we can take Eq.~\ref{eq:flrw metric 2}, invert it and calculate the comoving distance 
\begin{equation}
    d_c(t_s) = \int_0^{t_s} \frac{c dt}{a(t)}.
    \label{eq:dc}
\end{equation}
We can now recall the definition in Eq.~\ref{eq:defs}, and differentiate to obtain
\begin{eqnarray}
    dz &=& -\frac{\dot{a}(t)}{a^2(t)} dt \label{eq:cv} \\ 
    dt &=& \frac{dz}{H(z) (1+z)},
\end{eqnarray}
and rewrite Eq.~\ref{eq:dc} as 
\begin{equation}
    d_c(t_s) = \int_0^{z} \frac{c dz}{H(z)(1+z)} \approx \frac{c z}{H_0},
    \label{eq:dc2}
\end{equation}
where we have used the fact that very close to us (at redshifts $z \ll 1$), the Hubble parameter is just the Hubble constant $H_0$.  Eq.~\ref{eq:dc2} is yet another expression for the Hubble--Lemaitre law, where instead of the recessional velocity, we use the redshift of the source.
\end{itemize}

To summarize our discussion so far, in order to measure the cosmic expansion from individual sources, we need to measure the distance of the object and its recessional velocity (or redshift). For EM sources, the redshift can be obtained thanks to the measurement of the redshift of spectral emission lines of elements, while the distance is less trivial to obtain. 
To obtain distances for EM sources at cosmological distances, we need \textit{standard candles}, namely a class of sources for which we know the intrinsic luminosity $L$. A standard candle allows us to measure the flux $F$ and then obtain the distance by inverting its flux equation given by
\begin{equation}
    F = \frac{L}{4\pi d^2}.
    \label{eq:flux_def_naive}
\end{equation}
In a static Euclidean Universe, this would define the physical distance \( d_p \) seen earlier. However, in an expanding Universe, this relation is modified by two key effects. First, the energy of each photon is redshifted by a factor \( (1+z) \), and the rate at which photons are received is also reduced by \( (1+z) \), leading to a total flux suppression of \( (1+z)^2 \). The proper observed flux is then corrected such that
\begin{equation}
    F = \frac{L}{4\pi d_c^2 (1+z)^2}.
\end{equation}
Comparing this with the definition in Eq.~\eqref{eq:flux_def_naive}, we identify the \textit{luminosity distance} as
\begin{equation}
    d_L(z) = (1+z) d_c(z),
    \label{eq:luminosity_distance}
\end{equation}
which effectively absorbs the redshift dependency, and it is the distance that we measure from standard candles. Here we do not aim for a dedicated review of standard candles (we refer the reader to references in \cite{Moresco:2022phi} for more details), let us just note that standard candles have some intrinsic limitations. Their intrinsic luminosity needs to be calibrated on closer standard candles, i.e. they need a \textit{cosmological ladder}, and this can introduce systematics in their calibration. Moreover, only a very restricted ensemble of sources, such as Cepheid and Supernova Type IA, can be standardized.

The relationship between luminosity distance and redshift (comoving distance) in Eq.~\ref{eq:luminosity_distance} is the main ingredient required for \textit{standard candles} and \textit{standard sirens} cosmology. Now, we have seen that at low redshifts, $d_c$ can be expressed in terms of the Hubble constant; however, at higher redshifts, the scale factor and the universe expansion are dominated by other cosmological parameters.
At this point, we need to introduce the first Friedmann's 
equation\begin{equation}
    \left( \frac{H(z)}{H_0} \right)^2 = \Omega_{\rm m}(1+z)^3 + \Omega_{\Lambda},
    \label{eq:first friedmann bis}
\end{equation}
that connects the evolution of the Hubble parameter (and hence the scale factor) to the Hubble constant, the dark matter energy density $\Omega_{\rm m}$ and the dark energy density $\Omega_{\Lambda}$\footnote{Here, remember that we assume that the Universe is flat, i.e. $\Omega_{k}=0$.}. The first Friedmann's equation is a direct product of Einstein's field equations, supplied with the FRLW metric and a stress-energy tensor of an ideal fluid, where some polytropic relations are assumed for all the energy density components of the stress-energy tensor. We refer the reader to \cite{Dodelson:2003ft} for a complete derivation, here we just note that with Eq.~\ref{eq:first friedmann bis}, we describe a flat Universe dominated by dark matter and dark energy. When deriving the Friedmann equation under the assumption of an ideal fluid, one can define the critical density of the Universe as
\begin{equation}
    \rho_c = \frac{3H_0^2}{8 \pi G},
\end{equation}
which represents the energy density required for a spatially flat Universe. All density parameters are then expressed relative to this critical density.
Remembering the expression of the comoving distance in Eq.~\ref{eq:dc2} and Friedmann's Eq.~\ref{eq:first friedmann bis}, we can rewrite the full expression for the luminosity distance as
\begin{equation}
    d_L(z) =(1+z) d_c(z) =(1+z) \int_0^z \frac{dz'}{H_0 \sqrt{\Omega_{\rm m}(1+z')^3 + \Omega_{\Lambda}}}.
    \label{luminosity_distance_finale}
\end{equation}
Another relevant quantity that we will need for GW cosmology is the notion of differential of comoving volume $dV_c$. The comoving volume for a flat universe, given by
\begin{equation}
V_c = \frac{4}{3} \pi d_c^3(z),
\end{equation}
is often used to calculate the number density of compact objects, or even galaxies present in a certain volume. Its differential as a function of redshift is hence given by
\begin{equation}
    \frac{dV_c}{dz} = 4\pi \frac{c}{H_0}\frac{d_c^2(z)}{\sqrt{\Omega_{\rm m}(1+z')^3 + \Omega_{\Lambda}}}.
\end{equation}

\begin{figure}
    \centering
    \includegraphics[scale=1.0]{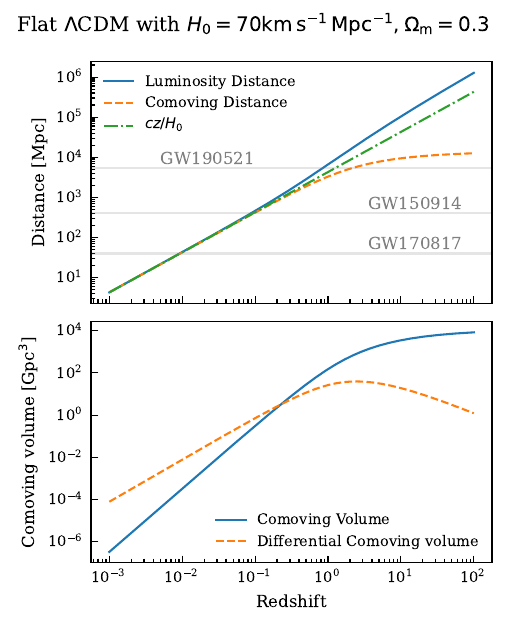}
    \caption{\textbf{Top panel:} Different definition of distances in cosmology compared to the median distance reported by three of the most discussed GW sources. \textbf{Bottom panel:} Comoving volume and its differential as a function of redshift. Both the panels have been generated using a flat $\Lambda$CDM model, with $\rm H_0=70$ $\rm kms^{-1}Mpc^{-1}$ and $\Omega_{\rm m}=0.3$.}
    \label{fig:dv}
\end{figure}
To conclude our introduction to standard cosmology, it is instructive to compare the typical distances and volumes relevant for GW observations. Figure~\ref{fig:dv} illustrates the various distance definitions introduced in this chapter for a flat $\Lambda$CDM Universe, along with the typical distances at which GW sources are observed. As we will discuss later, GWs allow for a direct measurement of the source's luminosity distance. For the most distant GW events, such as GW190521, it becomes necessary to use the full cosmological expression for the luminosity distance in terms of redshift and cosmological parameters.

\subsection{Compact binary coalescence at cosmological distance}
In this section we introduce the relevant ingredients required for GW cosmology with CBCs. We refer the reader to the GW theory section of this book for a more in-depth introduction about GWs and their emission from CBCs.

\begin{figure}
    \centering
    \includegraphics[width=0.8\textwidth]{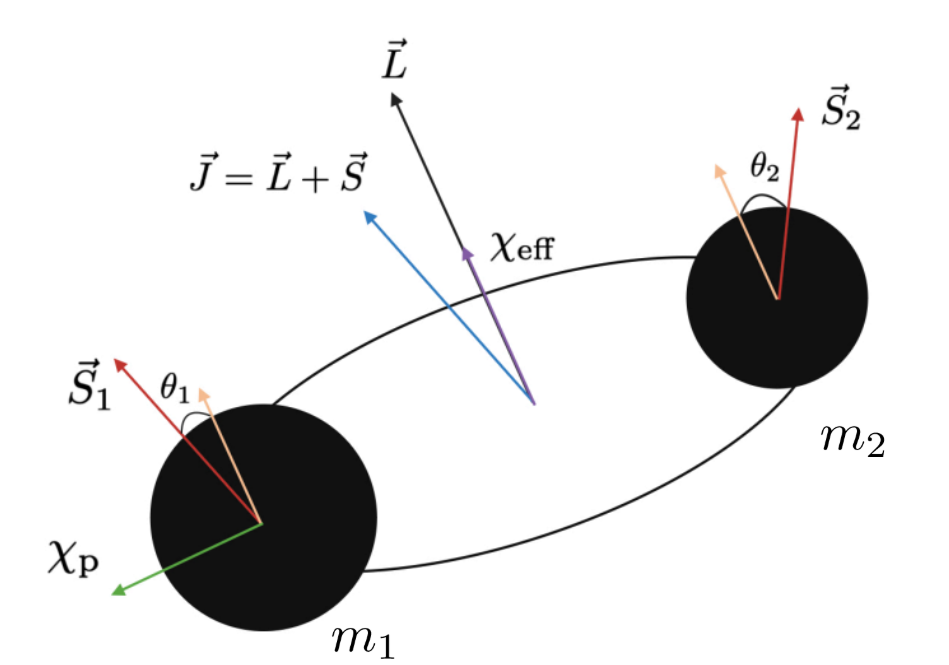}
    \caption{Schematic view of a binary system of compact objects with their main intrinsic parameters such as the primary and secondary masses $(\rm m_1,m_2)$, their spin vectors $(\rm S_{1},S_{2})$ and their inclination angles $(\theta_1,\theta_2)$.}
    \label{fig:bindraw.png}
\end{figure}

A CBC is characterized by a set of parameters that can be classified as either \textit{intrinsic} or \textit{extrinsic}. The intrinsic parameters describe properties that are inherent to the binary system, including the masses ($\rm m_{i}$), spins ($\chi_{i}$ and $\theta_{i}$) of the individual objects, their tidal deformability ($\Lambda_{i}$) if we are talking about NSs, and the orbital eccentricity. Usually, the two spins (and their orientations) are combined in two spin parameters; the effective spin parameter $\chi_{\rm eff}$ and the precession spin parameter $\chi_{\rm p}$. The former is maximized when the two spins are aligned to the orbital angular momentum, the latter when one of the two components lies on the orbital plane. 
A schematic representation of the binary intrinsic parameters is provided in Fig.~\ref{fig:bindraw.png}.
The extrinsic parameters depend instead on the observer's perspective. These include the binary luminosity distance (or redshift), orbital inclination relative to the line--of--sight ($\iota$), the time of merger as observed at the detector ($t_m$), the sky position (right ascension $\alpha$ and declination $\delta$), and the coalescence phase $\phi_c$ and polarization angle $\psi$. Together, these parameters define the key features of a CBC and play an essential role in interpreting GW signals, especially in the context of cosmological studies.

The GW strain observed at the detector, which refers to the spatial deformation caused by the passage of GWs through the interferometer, can be expressed as a combination of both intrinsic and extrinsic parameters in the detector frame. The strain at a given time, $t_m$, is given by the equation:
\begin{equation}
    h(t_m) = F_{+}(\alpha,\delta,\psi,t_m)h_{+}(t_m) + F_{\rm x}(\alpha,\delta,\psi,t_m)h_{\rm x}(t_m),
    \label{eq: gw strain}
\end{equation}
where the functions $F_{+}$ and $F_{\rm x}$ represent the detector\textquotesingle s response to the two GW polarizations, $h_{+}$ and $h_{\rm x}$. These response functions depend on several factors, including the geometry of the detector, the sky position of the source, the time of arrival of the signal, and the polarisation angle. For ground-based GW detectors, CBC signals typically last from a few minutes to a few milliseconds. As a result, the response functions $F_{+}$ and $F_{\rm x}$ can be considered constant over the duration of the signal, as their variation occurs over time scales comparable to the sidereal day.


In Fourier space, under the quadrupole approximation and assuming circular orbits, the stationary phase approximation, negligible spin effects, and at the leading order of the Post-Netwonian (PN) expansion, the two GW polarizations can be expressed as:  
\begin{equation}
    \tilde{h}_{+}(f) = \mathcal{A}(f, \mathcal{M}) \, \frac{1 + \cos^2 \iota}{2} \, e^{i \Psi(f, \mathcal{M})},
\end{equation}
and 
\begin{equation}
    \tilde{h}_{\times}(f) = \mathcal{A}(f, \mathcal{M}) \, \cos \iota \, e^{i\left( \frac{\pi}{2} + \Psi(f, \mathcal{M}) \right)},
\end{equation}
where the phase $\Psi(f,\mathcal{M})$ is expressed as
\begin{equation}
    \Psi(f, \mathcal{M}) = 2\pi f t_m - \frac{\pi}{4} - \phi_c + \frac{3}{128} \left( \frac{\pi G \mathcal{M}}{c^3} \right)^{-5/3} \frac{1}{f^{5/3}},
    \label{eq: GW phase}
\end{equation}
and the amplitude $\mathcal{A}(f,\mathcal{M})$ as
\begin{equation}
    A(f, \mathcal{M}) = \frac{1}{d_{p}} \frac{5}{24\pi^{4/3}} \frac{(G \mathcal{M})^{5/6}}{c^{3/2}} \frac{1}{f^{7/6}}.
    \label{eq: GW amp}
\end{equation}
In Eq.~\ref{eq: GW phase} and Eq.~\ref{eq: GW amp}, $f$ is the frequency of the GW and $d_p$ is the physical distance of the source to the detector and the chirp mass $\mathcal{M}$ is defined as
\begin{equation}
    \mathcal{M} = \frac{(m_1 m_2)^{3/5}}{(m_1 + m_2)^{1/5}}, 
\end{equation}
computed as a combination between the primary ($m_1$) and secondary ($m_2$) masses of the coalescing binary system. Usually, the convention between $m_1$ and $m_2$ is that the primary mass is larger or equal to the secondary so that $q=m_2 / m_1  \leq 1$. The chirp mass of the two objects is typically well-measured from the GW phase, while the distance is measured from the amplitude term.

As discussed previously, when we consider GW signals originating from the coalescences of binary systems located at cosmological distances, we have to account for the effect of the cosmological redshift due to the expansion of the Universe. From now on, we will use the subscript ``$\rm d$'' when we refer to detector quantities and ``$\rm s$'' when we refer to source quantities. 
The first aspect that we need to consider is that now the Fourier space must be computed at the detector, meaning that we should consider the cosmic expansion. 
This can be done easily using the properties of the Fourier transforms. At the detector, the signal in time domain can be calculated from the signal in time domain at the source, considering
\begin{equation}
    h_d(t_d) = h_s\left(\frac{t_d}{1+z}\right).
\end{equation}
We can now apply the Fourier trasform in detector time domain
 $\mathcal{F}_d$,
\begin{eqnarray}
    \mathcal{F}_d[h_d(t_d)] &=& \mathcal{F}_d\left[h_s\left(\frac{t_d}{1+z}\right)\right] \nonumber \\ &=& (1+z) \mathcal{F}_d[h_s(t_d)](f_d(1+z)) \nonumber \\ &=& (1+z) \tilde{h}_s(f_d(1+z)).
    \label{eq:eq33}
\end{eqnarray}
In the second step above, we have used the properties of the Fourier transforms. 
In writing Eq.~\ref{eq:eq33}, we have taken into account the fact that the frequency is redshifted. We have also dropped the polarization index to indicate that this equation is valid for both polarizations. We can see that in the detector, the signal is the original signal, multiplied by a redshift factor and calculated at frequencies that are the redshifted ones. Eq.~\ref{eq:eq33} is a standard relation in cosmology, in the limit that $z \rightarrow \infty$, the spectrum at the detector will be concentrated at low frequencies with a large amplitude.

Let us now explicitly expand Eq.~\ref{eq:eq33}. We will start with the phase term in Eq.~\ref{eq: GW phase}. If no additional phase will arise from the amplitude of the GW in Eq.~\ref{eq: GW amp} (as we will show later), then the phase of the GW signal at the detector is 
\begin{equation}
    \Psi_{\rm d}(f_d) = \Psi_{\rm s}(f_d/(1+z)). 
\end{equation}
Given the above equation, to obtain the phase at the detector, we simply need to replace $f_{\rm s}=f_{\rm d}(1+z)$ and also remember that the inverse is true for times (i.e $t_{\rm s}=t_{\rm d}/(1+z)$).
Eq.~\ref{eq: GW phase} can then be written as
\begin{eqnarray}
    \Psi_{\rm d}(f_d, \mathcal{M}_d) &=& 2\pi f_d(1+z) \frac{t_{m,d}}{1+z} - \frac{\pi}{4} - \phi_c + \frac{3}{128} \left( \frac{\pi G \mathcal{M}}{c^3} \right)^{-5/3} \frac{(1+z)^{-5/3}}{f_d^{5/3}} \nonumber\\
    &=& 2\pi f_d t_{m,d} - \frac{\pi}{4} - \phi_c + \frac{3}{128} \left( \frac{\pi G \mathcal{M}_d}{c^3} \right)^{-5/3}\frac{1}{f_d^{5/3}}.
    \label{eq: gw phase detframe}
\end{eqnarray}
The above equation expresses the GW source phase in terms of detector frequency and a newly defined quantity, the redshifted chirp mass $\mathcal{M}_d = (1+z)\mathcal{M}_s$. It is interesting to note that in the detector, the GW phase is mathematically equivalent to the phase of a GW signal with redshifted masses $\mathcal{M}_d = (1+z)\mathcal{M}_s$. 
Let us now see what happens to the GW amplitude in Eq.~\ref{eq: GW amp}. Again, we replace $f_{\rm s}=f_{\rm d}(1+z)$, and again we redefine the redshifted chirp mass $\mathcal{M}_{\rm s}=\mathcal{M}_{\rm d}/(1+z)$. With these definitions, we have
\begin{equation}
    \mathcal{A}(f_d,\mathcal{M}_d) = \frac{1}{d_{p}(1+z)^2} \frac{5}{24\pi^{4/3}} \frac{(G \mathcal{M}_d)^{5/6}}{c^{3/2}} \frac{1}{f_d^{7/6}}.
    \label{eq:gw amp det}
\end{equation}
We can now write explicitly Eq.~\ref{eq:eq33} using Eq.~\ref{eq:gw amp det} and Eq.~\ref{eq: gw phase detframe}. In doing so, we will remember that the luminosity distance is defined as $d_L=d_p(1+z)$, so that the two polarizations now become
\begin{eqnarray}
\tilde{h}_{d,+}(f_d) &=&  \frac{1}{d_{L}} \frac{5}{24\pi^{4/3}} \frac{(G \mathcal{M}_d)^{5/6}}{c^{3/2}} \frac{1}{f_d^{7/6}} \frac{1 + \cos^2 \iota}{2} \, e^{i \Psi(f_d, \mathcal{M}_d)} \label{eq:hpd}\\
\tilde{h}_{d,\times}(f_d) &=&  \frac{1}{d_{L}} \frac{5}{24\pi^{4/3}} \frac{(G \mathcal{M}_d)^{5/6}}{c^{3/2}} \frac{1}{f_d^{7/6}} \cos \iota \, e^{i\left( \frac{\pi}{2} + \Psi(f_d, \mathcal{M}_d) \right)} \label{eq:hpx}
\end{eqnarray}
These equations reveal several crucial information. 
\begin{itemize}
    \item The observed GW signal retains the same functional form as in the source frame, but with two important substitutions: the physical distance $d_p$ is replaced by the luminosity distance $d_L$, and the source-frame chirp mass $\mathcal{M}_s$ is replaced by the redshifted chirp mass $\mathcal{M}_d =(1+z)\mathcal{M}_s$.
    \item As a result, the waveform detected by ground-based interferometers provides two key observables: the redshifted chirp mass (from the phase evolution) and the luminosity distance (from the signal amplitude).
    \item However, these quantities are degenerate with redshift: one cannot disentangle the redshift from the mass based on the waveform alone. This fundamental limitation implies that the redshift of the source is not directly measurable from the GW signal, and additional information is needed to exploit standard sirens for cosmology.
\end{itemize}
Since for CBCs we can measure directly the distance via GW emission, we typically refer to this type of source as \textit{standard sirens}. Standard sirens are self-calibrated sources, as in contrast to EM sources, they do not require an intrinsic luminosity (calibration) to provide a distance measure.

\begin{figure}
    \centering
    \includegraphics[width=\textwidth]{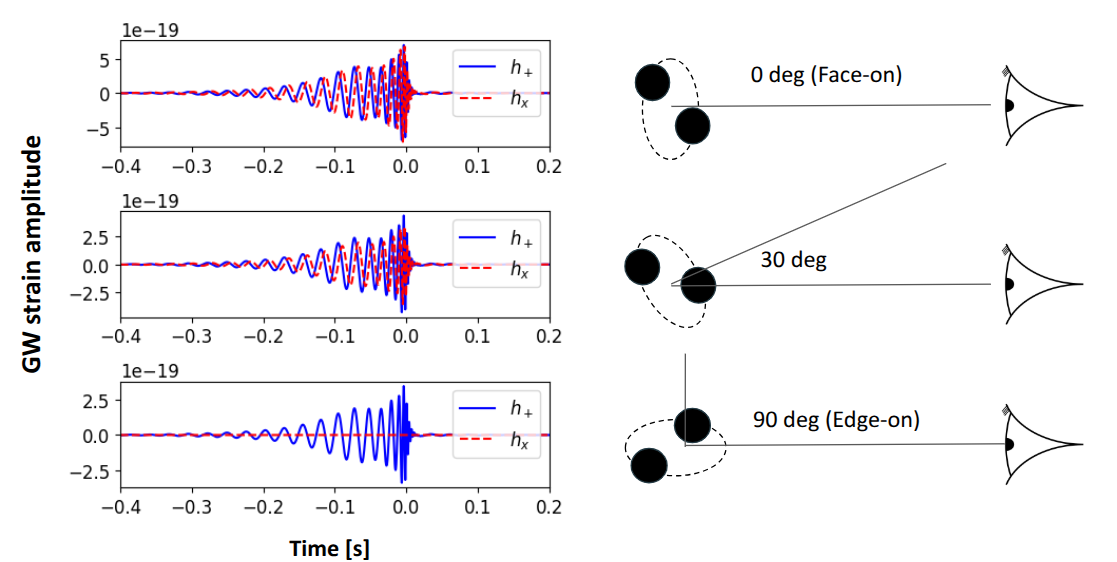}
    \caption{From top to bottom: Plus and cross GW emission for an equal-masses binary with redshifted chirp mass of 32 $M_\odot$ located at 500 Mpc. The figures on the right show the inclination of the binary with respect to the line of sight.}
    \label{fig:dldeg}
\end{figure}

An important feature of GW signals is the degeneracy between the signal amplitude and the inclination angle of the binary\textquotesingle s orbit relative to the observer. This degeneracy is the motivation for which standard sirens typically do not provide a precise measure of their distance.
Specifically, the GW amplitude depends strongly on $cos\iota$, see Eq.~\ref{eq:hpd}-\ref{eq:hpx}, making it difficult to distinguish between a nearby system viewed edge-on and a more distant system viewed face-on. 
A depiction of this effect from the point of view of the waveform at the detector is given in Fig.~\ref{fig:dldeg}. As we can see from the plots, the emission of GWs is stronger perpendicularly to the orbital plane (as both polarizations are maximized), while it is weaker parallel to the orbital plane. This effect greatly impacts the overall amplitude of the resulting signal, as the luminosity distance would.
This degeneracy limits our ability to precisely infer the luminosity distance $d_L$ from the observed signal. Fortunately, it can be mitigated by using waveform models that include additional physical effects, such as precession or higher-order modes. These effects imprint weak signatures in the signal that help to lift the degeneracy between inclination and distance, thereby improving our estimates of the luminosity distance. Accurate distance measurements are particularly important in GW cosmology, as they directly affect the precision with which we can constrain key cosmological parameters.

In summary, measuring the expansion of the universe requires both the luminosity distance $d_L$ and the redshift $z$ of each source. CBCs detected through GWs are the only known astrophysical sources that provide a direct measurement of the luminosity distance through the amplitude of their signal. From the waveform, we also extract detector-frame masses, but the redshift remains inaccessible due to its degeneracy with the source mass. Therefore, while GW sources have the potential to become powerful cosmological probes, additional information or assumptions are needed to determine $z$. The following sections review current approaches to assigning redshifts to GW sources.

\section{Bayesian inference for cosmological and population properties of resolved sources}
\label{sec:2}
In this section, we introduce the hierarchical Bayesian inference (HBI) framework, the main statistical tool currently used for standard sirens cosmology. We will start by introducing the basics of Bayesian statistics and then derive the hierarchical likelihood in the case of GW standard sirens.

\subsection{A brief recap of Bayesian theory}

The main scope of Bayesian inference is to estimate a probability density function for a certain parameter $\theta$ given the some observations $\{x\}$. Given a certain model $M$, the Bayes' theorem allows us to write this probability as
\begin{equation}
    P(\theta|\{x\}, M) = \frac{\mathcal{L}(\{x\}|\theta, M) \, P(\theta|M)}{P(\{x\}|M)},
    \label{eq:bayes}
\end{equation}
$\mathcal{L}(\{x\}|\theta, M)$ is the likelihood -- the probability of observing $\{x\}$ given $\theta$; $P(\theta|M)$ is the prior -- our knowledge about $\theta$ before seeing the data; $P(\{x\}|M)$ is the evidence -- a normalization factor used for model comparison; and $P(\theta|\{x\}, M)$ is the posterior -- the updated probability of $\theta$ given the data.
Here, we do not wish to go into the details of Bayesian statistics; we refer the reader to \cite{mackay2003information} for a more detailed introduction. We rather provide an operational example to understand its application.

\textbf{The mean of a gaussian:} Let's assume we are provided with a generator of random Gaussian numbers $x$. We are tasked to observe samples of $x$ and estimate the mean of the gaussian $\mu$. The first step is to write down the likelihood. The likelihood for obtaining the sample $x$  given the mean $\mu$ and standard deviation $\sigma$ of the normal distribution is 
\begin{equation}
  \mathcal{L}(x|\mu,\sigma)=\frac{1}{\sqrt{2\pi}\sigma}e^{-\frac{(x-\mu)^2}{2\sigma^2}}. 
\end{equation}
As different samples of $x$ are independent from each other, the overall likelihood can be written as 
\begin{equation}
    \mathcal{L}(\{x\}|\mu,\sigma)=\prod_{i=1}^N \mathcal{L}(x_i|\mu,\sigma)=\prod_{i=1}^N \frac{1}{\sqrt{2\pi}\sigma}e^{-\frac{(x_i-\mu)^2}{2\sigma^2}}.
\end{equation}
It can be shown that, if we define $\bar{\mu}=\sum_i x_i/N$ and $\bar{\sigma}=\sigma/\sqrt{N}$, then the above likelihood can be rewritten as  
\begin{equation}
    \mathcal{L}(\{x\}|\mu,\sigma)=\frac{1}{\sqrt{2\pi}\bar{\sigma}}e^{-\frac{(x_i-\bar{\mu})^2}{2\bar{\sigma}^2}}.
\end{equation}
We now use this likelihood in the Bayes' theorem (Eq.~\ref{eq:bayes}) and write explicitly the posterior
\begin{equation}
    p(\mu|\{x\},\sigma)=\frac{\mathcal{L}(\{x\}|\mu,\sigma) p(\mu)}{\int \mathcal{L}(\{x\}|\mu,\sigma) p(\mu) d\mu} \propto \frac{1}{\sqrt{2\pi}\bar{\sigma}}e^{-\frac{(x_i-\bar{\mu})^2}{2\bar{\sigma}^2}}.
    \label{eq:pos}
\end{equation}
In the last step, we have assumed a uniform prior on $\mu$ (constant) and neglected the denominator as it is just a normalization constant. We have also assumed to have perfect knowledge of $\sigma$ since we condition on it. We have discovered that the posterior on $\mu$ is itself a gaussian, centered around the mean of the dataset $\{x\}$ and with a standard deviation that becomes more and more precise as we have more observation ($\sqrt{N}$). This example helps us to understand the connection with frequentist statistics, where it is usually said that ``errors improve as $\sqrt{N}$''.
\begin{figure}
    \centering
    \includegraphics[scale=1.0]{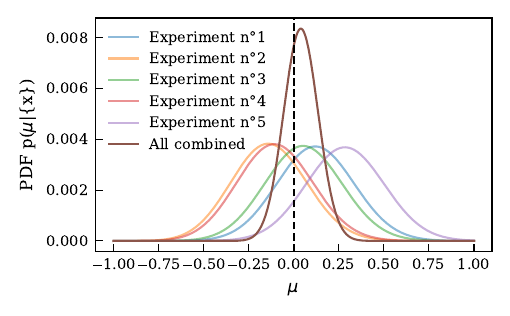}
    \caption{Posterior distributions (different colors) for the values of $\mu$ given 50 observations $x$ from a gaussian process. The vertical dashed line indicates the injected value of $\mu$.}
    \label{fig:measures_bayes}
\end{figure}

Our results can be easily validated numerically, we can, for instance, write a code that draws $50$ samples from a normal distribution ($\mu=0,\sigma=1$) and then loop on the values of possible $\mu$ in the prior to calculate Eq.~\ref{eq:pos}. Fig.~\ref{fig:measures_bayes} shows the result for this code, where multiple \textit{realizations} of the 50 samples have been generated. One important aspect to notice is that the posterior does not have to present its mode at the true value of $\mu$, since the measure of $\mu$ fluctuates due to the noise realization. Although, for a large number of observations, $N \rightarrow \infty$, we expect the posterior to converge to a Dirac-function centered on $\mu$.
These realizations, in virtue of their independence, can be later combined to obtain the posterior that would be obtained using all the 250 samples.

\textbf{Introducing the selection bias:} Let us make use of the previous example to also introduce the concept of selection bias. This time, our experiment is able to register only values of $x>0$ (this is a threshold). Then the likelihood $\mathcal{L}(x|\mu,\sigma)$ of observing a value of $x$ should be re-normalized by the number of $x$ that you can effectively see. In other words, the likelihood of observing a data point $x$ should effectively be normalized considering all the possible values of $x$ that you can detect. This can be easily done as follows
\begin{equation}
  \mathcal{L}(x|\mu,\sigma) =\frac{e^{[-(x-\mu)^2/(2\sigma^2)]}}{\int^{\infty}_{x_{thr}} e^{[-(x-\mu)^2/(2\sigma^2)]} dx} = \frac{e^{[-(x-\mu)^2/(2\sigma^2)]}}{I(\mu,x_{thr})}  
\end{equation}
Because the range of samples is restricted, the normalization at the denominator is equal to an unknown constant called $I(\mu,x_{thr})$. So the denominator has become a function of $\mu$. 
The denominator approaches the value of the usual normalization of the gaussian distribution when $\mu \gg x_{\rm thr}$, meaning that we have a process for which we can see all the possible realizations, otherwise, it is smaller.
\begin{figure}
    \centering
    \includegraphics[width=0.48\linewidth]{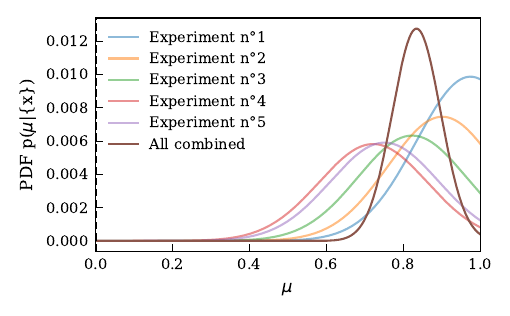}
    \includegraphics[width=0.48\linewidth]{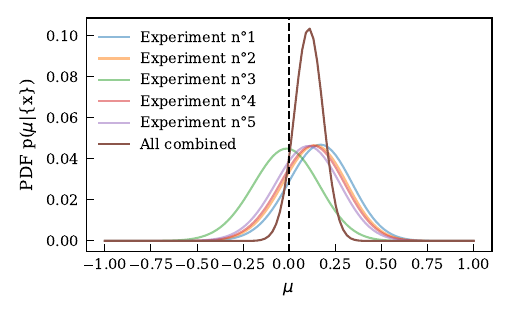}
    \caption{\textbf{Left:} Posterior distributions of the estimated value of $\mu$ with 50 observations of $x$ from a gaussian process without accounting he selection bias. The vertical dashed line indicates the injected value of $\mu$. \textbf{Right:} Same, but this time accounting for the selection bias.}
    \label{fig:biased}
\end{figure}
Fig.~\ref{fig:biased} shows us the difference of a Bayesian inference with and without selection bias for this case. When the selection bias is not accounted for, we would find a posterior supporting higher values of $\mu$. This is the consequence of the fact that we are not inferring the inference that observing values with $x<0$ is forbidden.

\subsection{Hierarchical Bayesian Inference with gravitational waves}

We have seen that the first step to using Bayesian statistics is to write a likelihood. Different from the previous case, where we wanted to estimate a parameter $\mu$ common to all the observations, now we have an extra level of inference. First, we need to infer the binary parameters for each observation (which are different among sources), then we want to infer population-level parameters such as the Hubble constant. Because of that, this kind of inference is called hierarchical, meaning we have a hierarchy of parameters. We are now going to construct the hierarchical likelihood for that scenario, following \cite{Mandel:2018mve}, and a more motivated reader can also follow a more accurate and detailed derivation in \cite{Vitale:2020aaz}.

We start our derivation by modeling the probability of generating a binary with parameters $\theta$ given a set of population parameters $\Lambda$. 
\begin{equation}
    p_{\rm pop}(\theta|\Lambda) = \frac{T}{N} \frac{dN}{d\theta dt_d}
    \label{eq:rate}
\end{equation}
For this derivation, $N$ is the total number of sources in the universe, $T$ the observing time at the detector and $\frac{dN}{d\theta dt_d}$ the differential number of sources (rate) per binary parameter (at the detector) and detector time. In writing the above identity, we have also assumed that the rate is constant in detector time (a reasonable assumption for cosmological sources as their rate do not change on human timescales), and the factor $T$ is basically the result of the integration in $dt_d$. It follows that the probability of having a population of GW sources with parameters $\{\theta\}$ is 
\begin{equation}
    p_{\rm pop}(\{\theta\}|\Lambda) = \prod_i^N \frac{p_{\rm pop}(\theta_i|\Lambda)}{\int d\theta_i p_{\rm pop}(\theta_i|\Lambda)}.
\end{equation}
Now, let us imagine that we again have the presence of a selection bias (finite detector sensitivity), as for the case pf the gaussian example, then we should also normalize the likelihood accordingly
\begin{equation}
    p_{\rm d}(\{\theta\}|\Lambda) = \prod_i^N \frac{p_{\rm pop}(\theta_i|\Lambda)}{\int d\theta p_{\rm pop}(\theta|\Lambda)p(\rm DET=1|\theta)} ,
    \label{eq:pdl}
\end{equation}
where the detection probability
\begin{equation}
    p(\rm DET=1|\theta) = \int_{x \in {\rm DET}} \mathcal{L}(x|\theta) dx,
\end{equation}
is the integral of the GW likelihood over all the possible realizations of data that are detectable.
If we would have been able to measure \textit{perfectly} the binary parameters $\theta$, we could have used directly \ref{eq:pdl} for our inference as likelihood. However, we do not have the luxury to perfectly measure $\theta_i$, as we also have a GW likelihood $\mathcal{L}(x_i|\theta)$. As such, what we measure for the single event is actually 
\begin{equation}
    p(x_i|\Lambda) = \int d\theta \mathcal{L}(x_i|\theta) p_{\rm d}(\theta|\Lambda) = \frac{\int d\theta \mathcal{L}(x_i|\theta) p_{\rm pop}(\theta|\Lambda)}{\int d\theta p_{\rm pop}(\theta|\Lambda)p(\rm DET=1|\theta)}.
\end{equation}
Finally, as all the GW events are independent from each other, we can write the hierarchical likelihood
\begin{equation}
    \mathcal{L}(\{x\}|\Lambda) = \prod_i^N p(x_i|\Lambda) = \prod_i^N \frac{\int d\theta \mathcal{L}(x_i|\theta) p_{\rm pop}(\theta_i|\Lambda)}{\int d\theta p_{\rm pop}(\theta|\Lambda)p(\rm DET=1|\theta)}.
    \label{eq:hbi_sf}
\end{equation}
The hierarchical likelihood can also be written in terms of physical rates. To do so, we need to introduce the expected number of detections
\begin{equation}
    N_{\rm exp}(\Lambda)= T \int d\theta \frac{dN}{d\theta t_d} p(\rm DET=1|\theta),
\end{equation}
and add a Poisson term to the likelihood
\begin{eqnarray}
    \mathcal{L}(\{x\}|\Lambda,N_{\rm exp}) \propto e^{-N_{\rm exp}(\Lambda)}[N_{\rm exp}(\Lambda)]^N \prod_i^N \frac{\int d\theta \mathcal{L}(x_i|\theta) p_{\rm pop}(\theta_i|\Lambda)}{\int d\theta p_{\rm pop}(\theta|\Lambda)p(\rm DET=1|\theta)} \nonumber\\.
    \label{eq:hbi}
\end{eqnarray}
Using the rate parameterization, the hierarchical likelihood in Eq.~\ref{eq:rate} can be expressed as
\begin{eqnarray}
    \boxed{
    \mathcal{L}(\{x\}|\Lambda,N_{\rm exp}) \propto e^{-N_{\rm exp}(\Lambda)} \prod_i^N T \int d\theta \mathcal{L}(x_i|\theta) \frac{dN}{d\theta dt_d}(\Lambda).}
    \label{eq:hbi2}
\end{eqnarray}
There also exist an equivalent version of that likelihood called ``scale-free'' likelihood. Ones can marginalize analytically the $N_{\rm exp}$ term using a scale-free prior $\pi(N_{\rm exp}) \propto 1/N_{\rm exp}$ to obtain a formally equivalent expression to Eq.~\ref{eq:hbi2}.
\begin{eqnarray}
    \boxed{
    \mathcal{L}(\{x\}|\Lambda) \propto\prod_i^N \frac{\int d\theta \mathcal{L}(x_i|\theta) \frac{dN}{d\theta dt_d}(\Lambda)}{\int d\theta p(\rm DET=1|\theta) \frac{dN}{d\theta dt_d}(\Lambda)}.}
    \label{eq:hbi_sf2}
\end{eqnarray}
Eqs.~\ref{eq:hbi2}-\ref{eq:hbi_sf2} are the hierarchical likelihood used for GW cosmology.

In the context of GW cosmology, the binary parameters must be written in terms of \textit{detector} binary parameters, namely luminosity distance, detector frame masses and spin parameters. The information on cosmological parameters enters either from an EM external data through the redshift (see later), or by modeling the CBC merger rate in terms of redshift and source masses. In the latter case, the cosmological information comes from the fact that the rate carries information about the source masses and redshift, while the GW likelihood measures the redshifted masses and luminosity distance.
We can express the detector-frame merger rate as a function of the source-frame parameters through a change of variables:
\begin{eqnarray}
    \frac{\rm d N}{\rm dt_{d} d\theta} &=& \frac{\rm d N}{\rm dt_{s} d\theta_{s}}\frac{\rm d t_s}{\rm d t_d}\frac{1}{\rm det J_{\rm d \rightarrow s}} \\
    &=&\frac{\rm d N}{\rm dt_{s} d\theta_{s}}\frac{1}{1+z}\frac{1}{\rm det J_{\rm d \rightarrow s}}
    \label{eq:rate_det},
\end{eqnarray}
where the time differential transforms as $1/(1+z)$, and the Jacobian accounts for the change of variables. It is important to notice that 
Eqs.~\ref{eq:hbi2}-\ref{eq:hbi_sf2} can either be written in detector variables using and dealing with the Jacobian to calculate the source rate, or directly in source variables without having to deal with the Jacobian. The two are mathematically equivalent, but in the second case, we will need to explicitly remember that the GW likelihood and detection probability are conditioned on the cosmological parameters (as they depend on detector quantities).

Since only masses and distances are affected by the expansion of the Universe, the Jacobian for this transformation is given by
\begin{equation}
    \frac{1}{\rm det J_{\rm d \rightarrow s}} = \frac{\partial \rm d_L}{\partial z}(1+z)^2,
\end{equation}
and using Eq.~\ref{luminosity_distance_finale}, it can be further written as
\begin{equation}
    \frac{1}{\rm det J_{\rm d \rightarrow s}} = \left(\frac{\rm d_L(z)}{1+z} +c\frac{1+z}{H_0}\frac{1}{E(z)}\right)(1+z)^2.
\end{equation}
Thus, the detector-frame CBC merger rate becomes
\begin{equation}
    \frac{\rm d N}{\rm dt_{d} d\theta} = \frac{\rm d N_{CBC}}{\rm dt_{s} d\theta_{s}}\left(\frac{\rm d_L(z)}{1+z} +c\frac{1+z}{H_0}\frac{1}{E(z)}\right)(1+z).
\end{equation}
From this point, three different methodologies for population inference can be pursued, depending on the goals of the analysis and the available data. Each approach re-parameterizes the source-frame CBC rate accordingly.

\section{Spectral sirens: Cosmology with GW sources and their source mass spectrum}
\label{sec:4}

\begin{figure}
    \centering
    \includegraphics[width=1\linewidth]{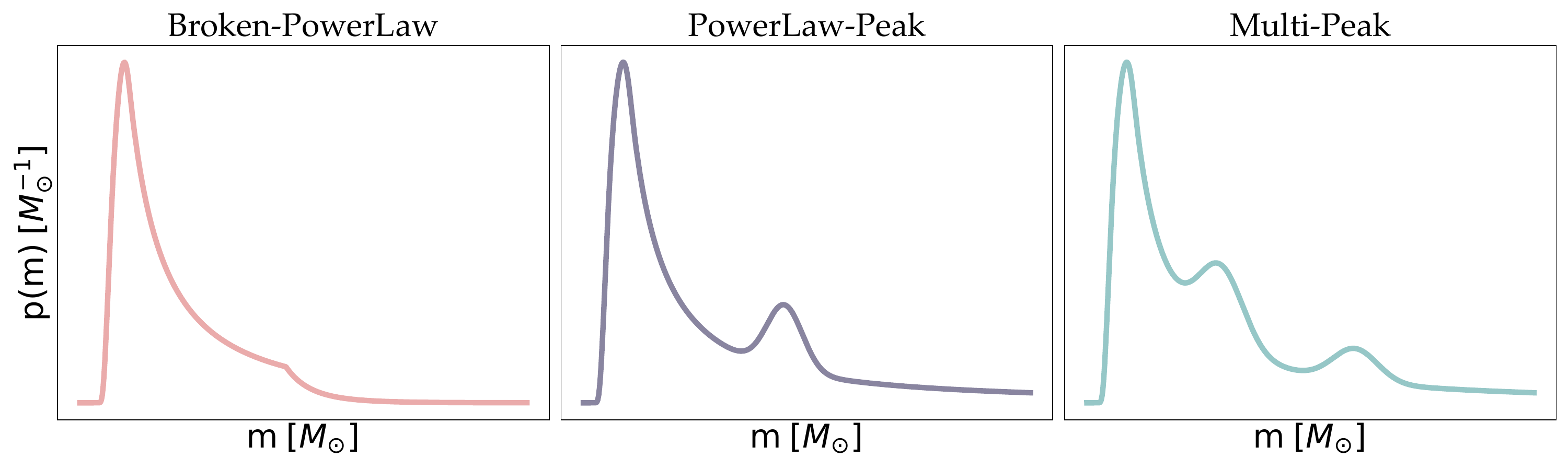}
    \caption{Representation of some mass models currently in vogue for spectral sirens cosmology to describe the mass spectrum of stellar-mass BHs, namely the $\rm Broken-PowerLaw$, the $\rm PowerLaw-Peak$ and the $\rm Multi-Peak$.}
    \label{fig:mass_model_schematic}
\end{figure}

One of the possibilities for obtaining redshift information from GWs alone is to make assumptions about the shape of the source-frame mass distribution of CBCs. The idea is the following: since the source and detector masses are related by a redshift factor of $(1+z)$, and since GW signals allow us to measure detector masses, we can statistically infer the redshift of each source by assuming a model for the source-frame mass distribution (with some free parameters).
This approach typically involves modeling the population distribution of CBCs in the source frame. In the following, let us explicitly separate the redshift $z$ from the rest of the source binary parameters $\theta_s$, we can write: 
\begin{equation} 
    \frac{\rm d N_{CBC}}{dt_{s} dz d\theta_{s}} = \frac{\rm d N_{CBC}}{dt_{s} dV_c d\theta_{s}}\frac{dV_c}{dz} =  \mathcal{R}_0\psi(z;\Lambda)p_{\rm pop}(\rm m_{1,s},m_{2,s}|\Lambda) \frac{dV_c}{dz}, 
    \label{eq: spectral siren rate} 
\end{equation} 
where $\mathcal{R}0$ is the local ($z=0$) CBC merger rate density per time and per comoving volume, typically expressed in $[\rm Gpc^{-3}yr^{-1}]$. The function $\psi(z;\Lambda)$ is a phenomenological model for the redshift evolution of the merger rate, usually inspired by the star formation rate. The term $p_{\rm pop}(\rm m_{1,s},m_{2,s}|\Lambda)$ represents the population model for the joint distribution of primary and secondary source masses. Note that Eq.~\ref{eq: spectral siren rate} already involves a key assumption: the mass distribution is independent of redshift. While this may not strictly hold in reality--since the population could evolve over cosmic time--current observations have not revealed any significant redshift dependence. 
For mass distribution, one can assume flexible analytical models like the ones represented in Fig.~\ref{fig:mass_model_schematic}, whose parameters are fit alongside the cosmological parameters. Here, we do not enter into the details of the modeling of the mass distribution and we refer the reader to \cite{Palmese:2025zku} for a review with more in-depth discussion.

\begin{figure}
    \centering
    \includegraphics[width=0.7\linewidth]{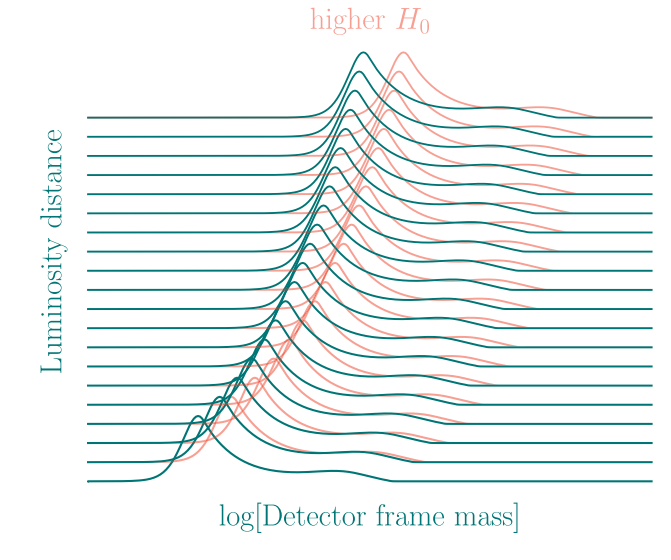}
    \caption{Detector mass distribution as a function of luminosity distance and the Hubble constant assuming a source mass model constructed with a powerlaw and a Gaussian peak, \textit{fixed} change in redshift. This figure is reproduced from \cite{Chen:2024gdn}.}
    \label{fig:spectral}
\end{figure}

To understand why the source mass distribution can provide information about cosmology, it is helpful to consider the impact of redshift on the observed masses. GW detectors measure masses in the detector frame, which are redshifted relative to the intrinsic source masses by a factor of $(1+z)$. If we had independent knowledge of the typical source masses of binary BHs (for instance, if we assumed they are all formed with primary masses around 35 $M_\odot$), then any deviation from this scale in the observed detector masses could be attributed to redshift. In this way, a characteristic mass scale in the source population effectively acts as a standard ruler in mass space, allowing the redshift to be inferred statistically. Then, with this redshift measure and the GW luminosity distance measure from the waveform, we can infer the cosmic expansion. 
Of course, we do not expect the binary source masses to have all the same values. Instead, we expect a distribution of source masses like the ones in Fig.~\ref{fig:mass_model_schematic}. Through the rate Eq.~\ref{eq: spectral siren rate}, this distribution is converted to a $d_L$-dependent detector mass distribution as displayed in Fig.~\ref{fig:spectral}. 
For a different combination of cosmological parameters $H_0,\Omega_m$, the resulting distribution of detector masses and function of $d_L$ will change because the luminosity distance is a function of $z$ and the cosmological parameters. HBI evaluates the fitting factor (the hierarchical) likelihood for this detector mass, luminosity distance distribution with real observed GW events.

\textbf{An example with real GW data:} It is interesting to provide here a simplified example of HBI in action for spectral sirens with current GW data. We will refer the reader to \cite{Palmese:2025zku} for a more complete description of the results obtained from data in the current literature.

To evaluate the hierarchical likelihood in Eq.~\ref{eq:hbi}, we mainly need two ingredients. The integrals of the rate models over the GW likelihood and the evaluation of the expected number of detections.
The former can be obtained via Monte Carlo integration by summing over posterior samples of the GW binary parameters, namely: 
\begin{eqnarray}
     \int \mathcal{L}(x_i|\theta) \frac{\rm d N}{dt_d d\theta}(\Lambda) d\theta &\approx& \frac{1}{d N_{{\rm s},i}} \sum_{j=1}^{N_{{\rm s},i}} \frac{1}{\pi_{\rm PE}(\theta_{i,j}|\Lambda)}\frac{dN}{dt_d d\theta}(\Lambda)\bigg|_{i,j}\\
     &\equiv& \frac{1}{N_{{\rm s},i}} \sum_{j=1}^{N_{{\rm s},i}} w_{i,j},
     \label{eq:intpe}
\end{eqnarray}
where the index $i$ refers to the event and the index $j$ to the posterior samples of the events. We have also defined a weight $w_{i,j}$ of dimension equal to the number of events generated per unit of time. 
The latter can be evaluated as: 
\begin{equation}
    N_{\rm exp}(\Lambda) = T \int p({\rm DET}=1|\theta) \frac{dN}{d t_d \theta} d \theta. 
    \label{eq:nexp}
\end{equation}
Typically, we do not have access to an analytical form of the detection probability (see \citet{Gair:2022zsa} for an introductory example in the context of GW cosmology with galaxy catalogs).
The current approach to evaluate selection biases is to use Monte Carlo simulations of injected and detected events, often shortly referred to as \textit{injections}. The injections are used to evaluate the volume that can be explored in the parameter space and correct for selection biases. Therefore, their occurrence is proportional to $p({\rm DET}=1|\theta)$ and the population model used to generate them.
Eq.~\eqref{eq:nexp} can also be approximated using Monte Carlo integration:
\begin{equation}
    N_{\rm exp}  \approx \frac{T}{N_{\rm gen}} \sum_{j=1}^{N_{\rm det}} \frac{1}{\pi_{\rm inj}(\theta_j)}\frac{dN}{dt_d d\theta}\bigg|_j \equiv \frac{T}{N_{\rm gen}} \sum_{j=1}^{N_{\rm det}} s_j=R_0 {\rm <VT>}.
    \label{eq:nexpnum}
\end{equation}
Here we have again defined a weight $s_j$ with the dimension of a rate of events. Note that there is one fundamental difference with Eq.~\eqref{eq:intpe}. The injection prior $\pi_{\rm inj}(\theta)$ must be properly normalized to obtain a reasonable value of $N_{\rm exp}$, while a wrong normalization of $\pi_{\rm PE}(\theta)$ (which is used in Eq.~\eqref{eq:intpe}) will only result in an overall normalization factor to the overall hierarchical likelihood.
Note that Eq.~\ref{eq:nexpnum} can also be used to define an interesting quantity, the explorable spacetime volume ${\rm <VT>}$ that quantifies how many ${\rm Gpc}^3$ per year we are able to explore with our GW detectors. The explorable space-time volume can be obtained as
\begin{eqnarray}
    {\rm <VT>} &\equiv& \frac{T}{R_0 N_{\rm gen}} \sum_{j=1}^{N_{\rm det}} s_j \\&=& \frac{T}{N_{\rm gen}} \sum_{j=1}^{N_{\rm det}} \psi(z^j;\Lambda)p_{\rm pop}(\rm m^j_{1,s},m^j_{2,s}|\Lambda) \frac{1}{1+z_i}\frac{1}{\rm det J_{\rm d \rightarrow s}|_i} \frac{dV_c}{dz}\bigg|_i, \nonumber \\
    \label{eq:VT}
\end{eqnarray}
where we have factorised the CBC merger rate at the detector using Eq.~\ref{eq: spectral siren rate}. Of course, coding up all the machinery required to perform HBI is not trivial, as it requires the implementation of cosmological and population models. For these lectures, however, we refer the reader to one of the python packages available for HBI with GWs, \textsc{icarogw} \citep{Mastrogiovanni:2023zbw} and its dedicated tutorials \citep{mastrogiovanni_2023_10135401}.
\begin{figure}
    \centering
    \includegraphics[width=0.48\linewidth]{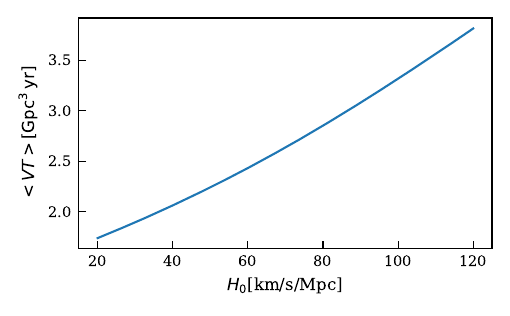}
    \includegraphics[width=0.48\linewidth]{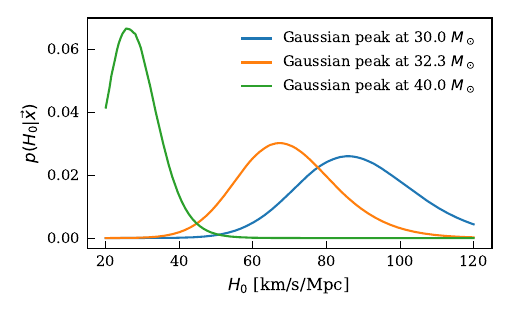}
    \caption{\textbf{Left:} Explorable spacetime volume for BBH signals with SNR>11 during the third LIGO--Virgo--KAGRA observing run (O3), as function of $H_0$. \textbf{Right:} Posterior distributions on $H_0$ calculated from 42 BBH signals from GWTC-3 with SNR>11, and using three different population models.}
    \label{fig:VT}
\end{figure}

Fig.~\ref{fig:VT} shows two of the relevant quantities for HBI that can be computed with \textsc{icarogw}. On the left side of the figure, we see the average spacetime volume as a function of $H_0$, while on the right, the posterior on $H_0$ for different choices of the mass distribution parameters. These plots have been generated by choosing a rate $\psi(z;\Lambda)$ that increases in redshift a power law plus peak model for the source masses \footnote{the reader can check the tutorials for the actual values of the population parameters}. The plots use a set of 42 BBHs with signal-to-noise ratio higher than 12 from the third GW transient catalog (GWTC-3) \cite{KAGRA:2021vkt}.
As shown in the plots, the average detectable space--time volume increases with the Hubble constant. This scaling does not arise solely from the ability to detect events at higher redshifts--although it is true that the redshift detection horizon scales as $z_{\rm hor} \propto H_0$--but also because the differential comoving volume scales as $\propto H_0^{-3}$. The increase in detectable volume is further amplified by the fact that higher values of the Hubble constant correspond to probing regions of the Universe where the CBC merger rate is higher, as encoded in $\psi(z; \Lambda)$. 

The right--hand plot illustrates another crucial point: population assumptions, particularly the mass model, significantly impact spectral siren cosmology. The displayed posteriors differ because they were generated using mass models with different Gaussian peak locations. This highlights the necessity of marginalizing over population parameters--such as those describing the mass spectrum--alongside cosmological parameters in current GW cosmology

\section{Adding galaxy surveys to dark sirens}
\label{sec:5}

\begin{figure}
    \centering
    \includegraphics[scale=0.75]{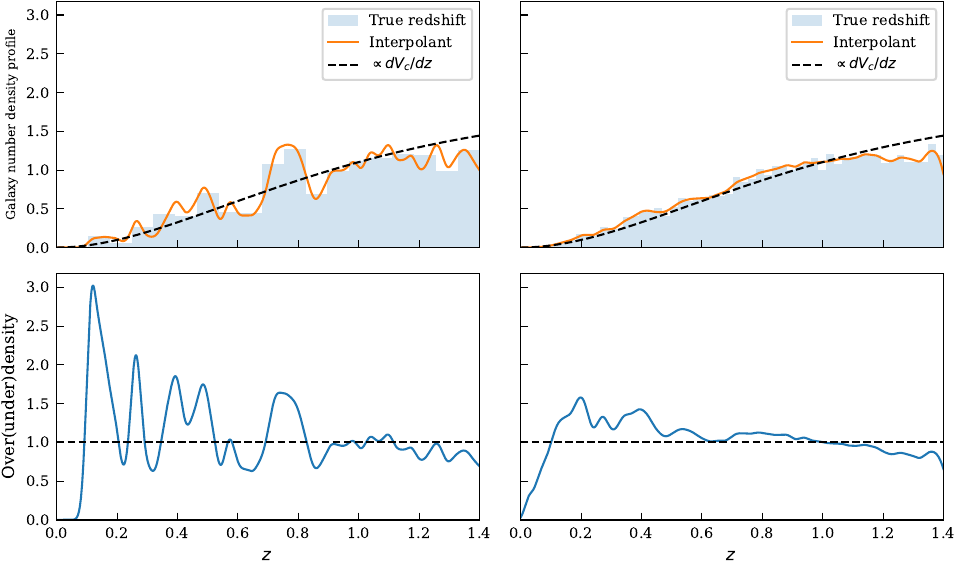}
    \caption{\textbf{Top plots:} Galaxy number density profile as a function of redshift from the simulated galaxy catalog used in \cite{Gair:2022zsa} for two random lines of sight. The distributions have been normalized in the corresponding redshift range. The black dashed line indicates a uniform in comoving volume redshift distribution. \textbf{Bottom plots:} Galaxy over/under density defined as the ratio between the galaxy number density profile and the uniform in comoving volume distribution.}
    \label{fig:pz}
\end{figure}

The first methodology that was actually proposed by \cite{1986Natur.323..310S} to use dark sirens for cosmology relates to the use of galaxy surveys. This idea involves using redshift information from galaxy surveys reported in the localization area of the GW event to identify potential hosts for each CBC event. In practice, what one can do is to build a redshift galaxy density profile for all the possible directions in which the source is localized, as shows in Fig.~\ref{fig:pz}. The galaxy redshift density profile is not really uniform in comoving volume, especially at low redshifts, and can present over densities and under densities. The most probable redshift for the GW is on the over densities.
Fig.~\ref{fig:pz} displays the logic behind this methodology.
It is interesting to note that, from a historical point of view, this was the first method proposed for cosmology with GW dark sirens. Nowadays, we know that the spectral sirens method (cosmology with the source mass) and the galaxy catalog method are actually part of the same method, in fact, for both the methods it is necessary to model the CBC merger rate.

In this textbook, we will use the ``rate'' approach to see how to implement a galaxy survey for GW cosmology. We will refer the reader to \cite{Gair:2022zsa} for a more complete step-by-step mathematical introduction on this topic from the point of view of probabilities. 
To parameterized the CBC merger rate as a function of the galaxy catalog, we can make two assumptions: first, that all CBC mergers occur within galaxies, and second, that the number of mergers per galaxy can be proportional to some galactic property, for instance the absolute magnitude $M$ and redshift. The CBC merger rate, now expressed in terms of the number of mergers per galaxy, is therefore rewritten as:
\begin{eqnarray} 
    \frac{\rm d N_{\rm CBC}}{\rm d t_s dz d \theta_{s}d\vec{\Omega}} &=& \int \rm dM\frac{\rm d N_{\rm CBC}}{\rm d t_s dz d \vec{m}_s d d\vec{\Omega} dM} \\ &=& \int \rm dM \frac{\rm d N_{\rm CBC}}{\rm d N_{\rm gal}d\vec{m}_s d t_s}\frac{\rm d N{\rm gal}}{\rm dz d\vec{\Omega} dM}, 
    \label{eq: cbc merger rate galaxy cat}
\end{eqnarray}
where $\rm N_{\rm gal}$ is the number of CBCs per galaxy, and $\vec{\Omega}=(\alpha,\delta)$ denotes the sky localization of the GW events.
This new parameterization of the CBC merger rate can be understood as the product of two terms. The first one closely resembles the expression used in classic spectral sirens and can be written as:
\begin{equation}
    \frac{\rm d N_{\rm CBC}}{\rm d N_{\rm gal}d\vec{m}_s d\vec{\chi}d t_s} = \mathcal{R}^{*}_{\rm gal,0}\Psi(z,M;\Lambda)p_{\rm pop}(\vec{m}_s|z,M;\Lambda).
    \label{eq: spectral term for the galaxy rate}
\end{equation}
Here, $\mathcal{R}^{*}_{\rm gal,0}$ replaces the usual local merger rate density and now represents the local CBC merger rate per galaxy per year. The function $\Psi(z,M;\Lambda)$ models how the rate of CBCs per galaxy evolves with redshift and magnitude, while the two population terms describe the source mass and spin distributions. It is a custom choice to model 
\begin{equation}
    \Psi(z,M;\Lambda) = \psi(z;\Lambda)10^{0.4\epsilon(M_*-M)},
\end{equation}
where $\psi(z;\Lambda)$ is the usual merger rate as function of redshift used for spectral sirens and the second term simply introduces a GW hosting probability proportional to the galaxy intrinsic luminosity, namely $\propto L^\epsilon$. All of these term rates, besides the one dependent on $M$, can be effectively taken out the integral in Eq.~\ref{eq: cbc merger rate galaxy cat}.

The second term in Eq.~\ref{eq: cbc merger rate galaxy cat}, is the galaxy number density per redshift, steradian and absolute magnitude. If we were supplied with a galaxy catalog containing all the galaxies in the universe, this term would simply be a ``histogram'' in redshift, sky position and absolute magnitude of all the galaxies. Instead, realistic galaxy catalogs are flux-limited, meaning that they only report galaxies brighter than a certain apparent magnitude threshold $m_{\rm thr}$. It follows that we need to apply a completeness correction by defining
\begin{equation}
    \frac{\rm d N{\rm gal}}{\rm dz d\vec{\Omega} dM} = \frac{\rm d N_{\rm gal, CAT}}{\rm dz d\vec{\Omega} dM} + \frac{\rm d N_{\rm gal, OUT}}{\rm dz d\vec{\Omega} dM},
    \label{eq:ciao}
\end{equation}
where $\rm d N_{\rm gal, CAT}$ and $\rm d N_{\rm gal, OUT}$ correspond to the number of galaxies reported by the galaxy catalog and brighter than $m_{\rm thr}$, and to the completeness correction.
The completeness correction is nonetheless given by the number density of missing galaxies in the redshift shell that can be computed as
\begin{equation}
    \frac{\rm d N_{\rm gal, OUT}}{\rm dz d\vec{\Omega} dM} = \Theta(M-M_{\rm thr}(m_{\rm thr},z,H_0)) {\rm Sch}(M) \frac{1}{4\pi} \frac{dV_c}{dz},
\end{equation}
where $\Theta$ is a Heaviside step that is non-null when the absolute magnitude is fainter than the threshold absolute magnitude, and ${\rm Sch}(M)$ is the Schecter function, namely a relation modelling the number density of galaxies in comoving volume per absolute magnitude.

Finally, the total CBC merger rate parameterization for the spectral siren analysis with a galaxy catalog becomes:
\begin{eqnarray}
    \frac{\rm d N_{\rm CBC}}{\rm d t_s dz d \theta_{s}d\vec{\Omega}} &=& \mathcal{R}^{*}_{\rm gal,0}p_{\rm pop}(m_{1,s},m_{2,s}|\Lambda) \psi(z;\Lambda) \times \nonumber \\  && \int dM 10^{0.4\epsilon(M_*-M)} \left( \frac{\rm d N_{\rm gal, CAT}}{\rm dz d\vec{\Omega} dM} + \frac{\rm d N_{\rm gal, OUT}}{\rm dz d\vec{\Omega} dM}\right).
    \label{eq: galaxy cat rate final}
\end{eqnarray}
Before moving to a real example with some data, let us comment on two limiting cases of Eq.~\ref{eq: galaxy cat rate final}. 
In the limit that the galaxy catalog is $100\%$ incomplete ($m_{\rm thr}=-\infty$), then the integral in Eq.~\ref{eq: galaxy cat rate final} is just the overall integral of the Schecter function (multiplied by the host probability). This integral can be computed analytically, and it results in Euler's gamma function, which acts as a normalization function. In this limiting case we see that (besides a normalizing factor), the rate is equivalent to the one of the spectral sirens. It follows that the mass distribution is fully dominating the implicit redshift information.
In the other limiting case, the galaxy catalog is 100\% complete and in Eq.~\ref{eq: galaxy cat rate final} only the catalog term contributes. In this case, if the galaxies are provided with perfect redshift and sky position measures, Eq.~\ref{eq: galaxy cat rate final} becomes 
\begin{eqnarray}
    \frac{\rm d N_{\rm CBC}}{\rm d t_s dz d \theta_{s}d\vec{\Omega}} &=& \mathcal{R}^{*}_{\rm gal,0}p_{\rm pop}(m_{1,s},m_{2,s}|\Lambda) \psi(z;\Lambda) \times \nonumber \\  && \sum_i^{\rm gal} \delta(\tilde{\Omega}-\Omega_i)\delta(z-z_i)10^{0.4\epsilon(M_*-M)}.
    \label{eq: galaxy cat rate final complete}
\end{eqnarray}
In the second scenario, the redshift and sky positions are entirely dictated by the galaxies, and the mass model acts as a weight factor between them. This shows us that the mass spectrum always enters the inference of galaxy catalogs.

\textbf{Galaxy catalogs in action:} As in the case of spectral sirens, the full inference of mass spectrum properties and the cosmological background parameters, with the addition of galaxy catalogs, is not an easy task. In this case, we have the additional difficulty of adding to the inference a galaxy survey that typically contains billions of galaxies that can make the computation of the hierarchical likelihood computationally inefficient. For this introductory text, we will follow \cite{Gair:2022zsa} and work under some main assumptions
\begin{itemize}
    \item We will neglect all the rate modeling associated to source masses and redshift of GW events (the function $\psi(z;\Lambda)$). We will further assume that GW events are at low redshifts and the $1/1+z$ factor coming from $dt_s/dt_d$ is not important.
    \item We will assume that the galaxy catalog is complete and it measures perfectly the galaxies.
    \item We will assume that galaxies are all equally likely to host GW events, whatever their properties.
\end{itemize}
Under these three main assumptions, that could be valid for very well localized and close GW events, the CBC merger rate in detector frame can be written as
\begin{eqnarray}
    \frac{\rm d N_{\rm CBC}}{\rm d t_d dz d\vec{\Omega}} &=& \mathcal{R}^{*}_{\rm gal,0} \sum_i^{\rm N_{\rm gal}} \delta(\tilde{\Omega}-\Omega_i)\delta(z-z_i).
\end{eqnarray}
If we note that the total number of CBC is $N_{\rm CBC} = \mathcal{R}^*_{\rm 0,gal} T N_{\rm gal}$ with $N_{\rm gal}$ the total number of galaxies in the universe, and recalling the definition for the population probability term in  Eq.~\ref{eq:rate}, we can also write that
\begin{equation}
    p_{\rm pop}(z,\tilde{\Omega}|\Lambda) = \frac{1}{N_{\rm gal}} \sum_i^{\rm N_{\rm gal}} \delta(\tilde{\Omega}-\Omega_i)\delta(z-z_i).
\end{equation}
Namely, we have obtained that the distribution of CBC sources can be described by a probability following the distribution of galaxies. We can now use the scale-free version of the hierarchical likelihood in Eq.~\ref{eq:hbi_sf} and analytically solve the integrals in $\tilde{\Omega}$ and $z$. In doing so, we will remember that the GW likelihood and detection probability actually depend on $d_L(z,H_0)$ and the sky direction, so that  
\begin{equation}
    \mathcal{L}(\{x\}|H_0) = \prod_i^{N_{\rm CBC}} \frac{\sum_{j}^{N_{\rm gal}} \mathcal{L}(x_i|d_L(z_j,H_0),\tilde{\Omega}_j)}{\sum_{j}^{N_{\rm gal}} p(\rm DET=1|d_L(z_j,H_0),\tilde{\Omega}_j)}.
    \label{eq:hbi_sf22}
\end{equation}
This equation is telling us that for well localized GW events, the hierarchical likelihood can be simply computed by evaluating the GW likelihood (or posterior in luminosity distance and sky location) with the reported redshifts and sky positions of galaxies. It is also interesting to note that the selection bias 
\begin{equation}
    \sum_{j}^{N_{\rm gal}} p(\rm DET=1|d_L(z_j,H_0),\tilde{\Omega}_j) \propto H_0^3,
\end{equation}
is proportional to the $H_0^3$. This is a consequence of the fact that the detection probability for GWs is a function of $d_L$ and as we change the Hubble constant, more and more galaxies enter inside the GW horizon. The number of galaxies that enter inside the luminosity distance horizon scales as a volume with redshift radius $z \propto H_0 d_{L,\rm horizon}$, and hence the $H_0^3$ scaling.

\begin{figure}
    \centering
    \includegraphics[width=1\linewidth]{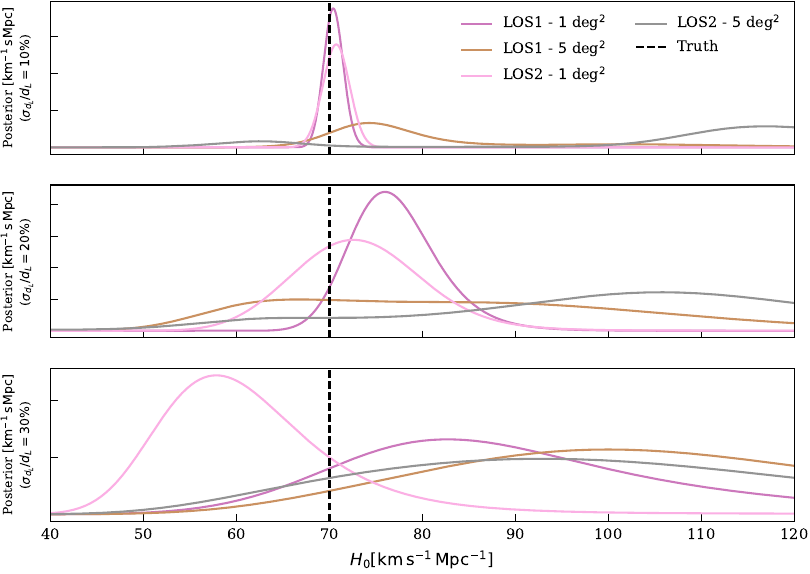}
    \caption{\textbf{From top to bottom} Hubble constant posteriors generated with 200 simulated GW events with varying sky error budget on $d_L$ and localized within $1$ deg$^2$ and $5$ deg$^2$  for two different lines of sight. The vertical dashed line marks the injected $H_0$ value. Posteriors are generated with codes released for \cite{Gair:2022zsa}.}
    \label{fig:big_fig_acc_redshift}
\end{figure}

In Fig.~\ref{fig:big_fig_acc_redshift}, we report a set of $H_0$ posteriors evaluated for Eq.~\ref{eq:hbi_sf22} with the codes\footnote{\url{https://github.com/simone-mastrogiovanni/hitchhiker_guide_dark_sirens}} released in \cite{Gair:2022zsa}.

\section{Bright sirens: The ``easiest'' case}
\label{sec:3}

\begin{figure}
    \centering
    \includegraphics[width=1.0\linewidth]{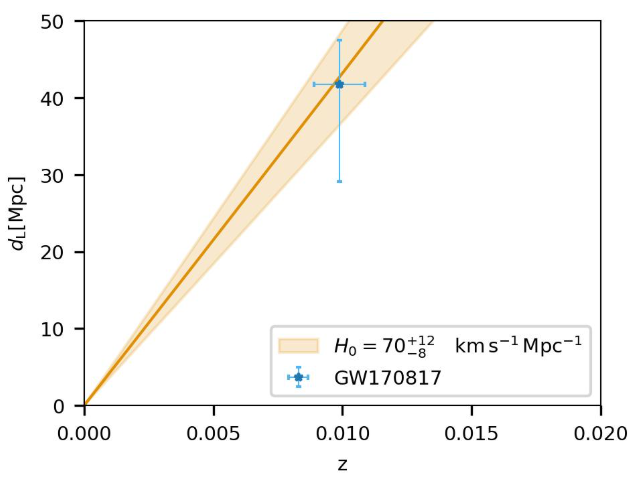}
    \caption{A $\chi^2$-like fit of the Hubble constant from the bright siren GW170817. The luminosity distance is inferred from the GW data and the redshift from the spectroscopic redshift of its host, NGC 4993. The orange contours is the $d_L(z)$ relation reconstructed assuming $d_(z)=cz/H_0$.}
    \label{fig:170817easy}
\end{figure}

We conclude our overview of methods for GW cosmology with the case for which an EM counterpart is observed jointly with the GW emission. This scenario, as demonstrated by GW170817 \cite{LIGOScientific:2017adf}, are extremely unlikely, due to the rarity of GW detection from BNSs and the difficulty in identifying the transient EM counterpart. Naively, one can use these bright sirens to rapidly estimate cosmological parameters as follows: the GW provides an estimation of the luminosity distance, the EM counterpart a redshift and then the cosmological parameters are fit with a relation linking $d_L(z)$ and minimizing a $\chi^2$ function. This is for instance what is reported in Fig.~\ref{fig:170817easy}.
However, while this treatment gives a rapid idea of the estimation of the cosmological parameters, it does not take into account the full complexity of selection biases at play with that type of detection. In this last section, we will see how to deal with EM counterparts within the hierarchical framework that we have discussed in the previously.

For bright sirens, the parameterization of the CBC merger rate is equivalent to that one of the spectral sirens in Eq.~\ref{eq: spectral siren rate}. The difference now is that the likelihood in Eq.~\ref{eq:hbi} is not anymore composed by solely the GW likelihood, but it is multiplied by the a new likelihood for the EM signal. From now on, let us work in source frame parameters, as it will be less tedious for calculations. The new likelihood can be written as follows
\begin{eqnarray}
    \mathcal{L}_{\rm GW + EM}(x_i|z,\vec{\Omega},\vec{\rm m}_s,\cos \iota)= && \mathcal{L}_{\rm GW}(x_i|d_L(z,H_0),\vec{\Omega},\vec{\rm m}(\vec{\rm m}_s,z),\cos \iota) \times \nonumber \\ && \mathcal{L}_{\rm EM}(x_i|z,\vec{\Omega},\cos \iota),
    \label{eq:em1}
\end{eqnarray}
where we explicitly have two terms, one likelihood for the GW part and one for the EM part. Eq.~\ref{eq:em1} tells us how well we measure the sky position, redshift, masses and inclination angle (and we will see why soon we included it back). At the same time, also the detection probability is now composed of a GW part and EM part,
\begin{eqnarray}
    p_{\rm GW + EM}({\rm DET=1}|z,\vec{\Omega},\vec{\rm m}_s,\cos \iota)= && p_{\rm GW}({\rm DET=1}|d_L(z,H_0),\vec{\Omega},\vec{\rm m}(\vec{\rm m}_s,z),\cos \iota) \times \nonumber \\ && p_{\rm EM}({\rm DET=1}|z,\vec{\Omega},\cos \iota). \nonumber \\
    \label{eq:em2}
\end{eqnarray}
Before continuing, let us inspect Eqs.\ref{eq:em1}-\ref{eq:em2}. In both equations, we have added back the contribution of $\cos \iota$, since if the EM counterpart is collimated with respect to the orbital plane (e.g. a  cone-shaped gamma ray burst), then the detection probability of the EM counterpart inherits a dependency on $\cos \iota$. In most of the current applications, it is assumed that the detection probability of an EM counterpart is more far reaching than the one of GW detections, meaning that $p_{\rm GW + EM}({\rm DET=1}|\ldots) \approx p_{\rm GW}({\rm DET=1}|\ldots)$. While this was a reasonable assumption for past GW observing runs, it is not anymore with the extended sensitivity range of GW detectors.
Moreover, it is possible to try to model the EM counterpart, for instance the gamma ray burst afterglow, to try to measure $\cos \iota$ from EM data from the likelihood $\mathcal{L}_{\rm EM}(x_i|z,\vec{\Omega},\cos \iota)$. Indeed, this is of crucial importance given the strong degeneracy on the GW side between $d_L$ and $\cos \iota$ (see Fig.~\ref{fig:dldeg}). 
With this digression, we just wanted to argue that the inclusion of the EM counterpart is extremely rewarding in terms of \textit{(i)} redshift localization and \textit{(ii)} breaking the luminosity distance, $\cos \iota$ degeneracy. However, it carries the burden of having to model the geometry of the EM counterpart and its detection probability.

Let us now derive the hierarchical likelihood under some simplified assumptions. We will assume that
\begin{itemize}
    \item There is no relevant information to obtain from the EM side on $\cos \iota$, hence that we can drop it from our equations.
    \item The GW detection probability dominates the EM detection probability, namely $p_{\rm GW + EM}({\rm DET=1}|\ldots) \approx p_{\rm GW}({\rm DET=1}|\ldots)$.
    \item The determination of masses from the GW likelihood, does not strongly correlate with the determination of the luminosity distance and sky position, namely
    \begin{eqnarray}
        \mathcal{L}_{\rm GW}(x_i|d_L(z,H_0),\vec{\Omega},\vec{\rm m}(\vec{\rm m}_s,z)) =&& \mathcal{L}_{\rm GW}(x_i|d_L(z,H_0),\vec{\Omega}) \times \nonumber \\  && \mathcal{L}_{\rm GW}(x_i|\vec{\rm m}(\vec{\rm m}_s,z)).
    \end{eqnarray}
    This is a reasonable assumption as masses are mostly dominated from the phase of the GW signal while the distance and sky position from the amplitude and arrival time at the detectors. It is also reasonable since the redshift of the EM counterpart is measured with a precision scale where the GW likelihood is typically not correlated in terms of luminosity distances and masses (it is like exposing the GW likelihood around a small interval in redshift).

    \item The EM counterpart is perfectly localized in redshift
    \begin{equation}
    \mathcal{L}_{\rm EM}(x_i|z,\vec{\Omega}) = \delta(z-z_i)\delta(\Omega-\Omega_i).
\end{equation}    
\end{itemize}
We now want to use Eq.~\ref{eq:em1}-\ref{eq:em2}, supplied with these assumptions to calculate Eq.~\ref{eq:hbi_sf2} using detectors variables $z,\vec{m}_s,\Omega$ and the rate parameterization in Eq.~\ref{eq: spectral siren rate}. 

\begin{figure}
    \centering
    \includegraphics[width=1.0\linewidth]{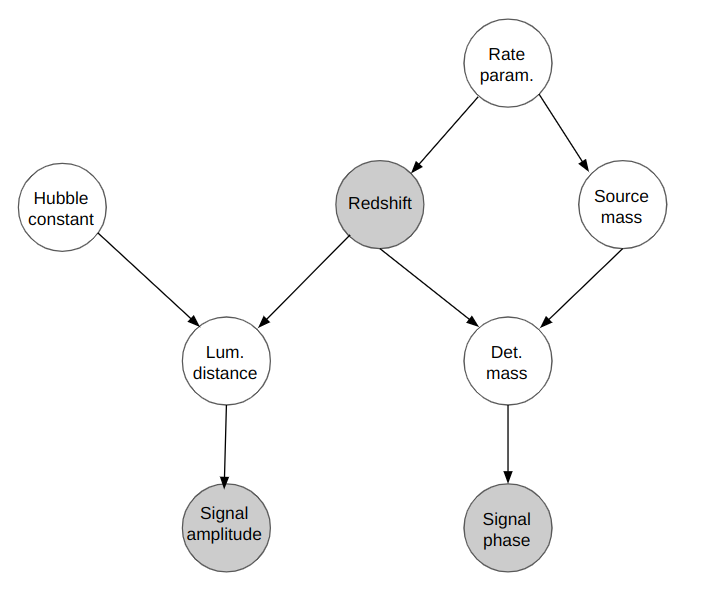}
    \caption{Schematic of a direct acyclic graph for the Hubble constant inference in a simplified bright siren case. The arrows represent conditional probabilities between nodes (random variables). The colored circle are observed quantities on which to condition. In this particular case, the redshift is perfectly measured and we estimate the luminosity distance and detector masses from amplitude and phase of the GW signal. According to the laws of probability, the posterior on $H_0$ given the redshift, detector, phase and amplitude is conditionally independent from the rest of the random variables to infer (including rate parameters)}
    \label{fig:dag}
\end{figure}

We start with the denominator, this is formally equivalent to what we would have for the spectral siren case, namely the expected number of detections in Eq.~\ref{eq:nexpnum}. It can also be noted that, as in the galaxy catalog case, if we neglect mass information the selection bias term scales as $H_0^3$ as more and more possible host galaxies enter in the GW detection horizon.
Considering a GW event $i$ with host galaxy $j$ and skipping some tedious steps for computation, we finally obtain that
\begin{eqnarray}
    I_{i,j}&=& \mathcal{L}_{\rm GW}(x_i|d_L(z_j,H_0),\vec{\Omega}_j) R_0 \frac{\psi(z_i|\Lambda)}{1+z_i} \frac{dV_c}{dz}\bigg|_{z=z_i} \times \nonumber \\ &&
    \int  \mathcal{L}_{\rm GW}(x_i|\vec{\rm m}(\vec{\rm m}_s,z_j)) p_{\rm pop}(\vec{\rm m}_s|\Lambda)d\vec{\rm m}_s.
\end{eqnarray}
In the above equation, everything besides $\mathcal{L}_{\rm GW}(x_i|d_L(z_j,H_0),\vec{\Omega}_j)$ acts as a normalization constant with respect to a varying $H_0$. This means that, even if the mass model that we apply is wrong (but still supports the values of masses found for the events), the inference on $H_0$ will still be unbiased. Statistically speaking, this is a consequence of the fact that, the determination of the GW luminosity distance from the signal amplitude and the masses from the signal phase are conditionally independent given the redshift. This means that no information related to source masses can propagate to $H_0$. Fig.~\ref{fig:dag} shows a depiction in terms of Bayesian direct acyclic graphs for this simplified bright siren scenario.

The hierarchical likelihood for the bright siren case can then be approximated as
\begin{eqnarray}
    \mathcal{L}(\{x\}|H_0,\Lambda) \propto \prod_i^N \frac{\mathcal{L}_{\rm GW}(x_i|d_L(z_j,H_0))}{H_0^3},
    \label{eq:lke}
\end{eqnarray}
assuming GW observatories sensitive only to low redshifts GW events. We warrant that for a real analysis, we would still need to consider possible uncertainties related to the population model.
If we want to include uncertainties on the redshift of the EM counterpart, we can still do it by using 
\begin{eqnarray}
    \mathcal{L}(\{x\}|H_0,\Lambda) \propto \prod_i^N \frac{\int dz \mathcal{L}_{\rm GW}(x_i|d_L(z_j,H_0)) \mathcal{L}_{\rm EM}(z_j|z)}{H_0^3},
    \label{eq:lke2}
\end{eqnarray}
where $\mathcal{L}_{\rm EM}(z_j|z)$ is a gaussian centered around the observed redshift. Eq.~\ref{eq:lke2} is valid in the limiting case that the redshift uncertainties from the EM side are much smaller than the typical redshift scale on which the CBC merger rate evolves so that we can still consider the rate as a normalization constant.

\section{Conclusions}

We conclude this chapter with a bit of historical discussion about GW cosmology. In 1929, Henrietta Leavitt discovered the first standard candles \cite{1912HarCi.173....1L}, the Cepheid, that were later used by Edwin Hubble \cite{1929PNAS...15..168H} to provide the first measure of the cosmic expansion (resulting in a biased $H_0=500$ km/s/Mpc). The error in the first Hubble's estimation of $H_0$ was later discovered to be a wrong calibration for galaxies distance. In the following 60 years, more standard candles were discovered, most notably Supernova type Ia, and new astrophysical calibrations for them were studied. Nowadays, the precision (and accuracy) on $H_0$ is of the order of a few per cent.

Almost 100 years later from the seminal discovery of Henrietta Leavitt, a new type of self-calibrating cosmological sources has been observed: GWs from compact binary coalescence. This exciting field started to be studied in 1986 by \cite{1986Natur.323..310S} and it has seen its dawn with the advent of the binary neutron star GW170817 in 2017, the first (and currently last) GW source with EM counterpart. Currently, GW cosmology is in the same situation as standard candles cosmology in the '50s. We have already explored the first GW sources to measure the cosmic expansion of the Universe and we are also beginning to learn how to properly calibrate the CBC merger rate to obtain an implicit redshift for dark sirens, in contrast to the luminosity distance of standard candles. As more and more GW detection are collected, the precision of the cosmological expansion parameters will improve, and in analogy to standard candles, we will also have to prove an excellent control of the systematics related to the calibration (statistical and astrophysical of our analyses. 

In summary,  these lectures only provides an introduction for GW cosmology, we expect more developments on the statistical and astrophysical aspects for GW cosmology will happen in the next years.

%
\end{document}